\documentclass[aps,prb,reprint,10pt,superscriptaddress,showpacs]{revtex4-1}

\usepackage{amsmath} 
\usepackage{amssymb} 
\usepackage{graphicx} 
\usepackage{hyperref} 
\usepackage{subfigure}

\begin{document}

\title{Modulation of Majorana-Induced Current Cross-Correlations by Quantum Dots}

\author{Bj\"orn Zocher}
\affiliation{Institut f\"ur Theoretische Physik, Universit\"at Leipzig, D-04103 Leipzig, Germany}
\affiliation{Max Planck Institut f\"ur Mathematik in den Naturwissenschaften, D-04103 Leipzig, Germany}

\author{Bernd Rosenow}
\affiliation{Institut f\"ur Theoretische Physik, Universit\"at Leipzig, D-04103 Leipzig, Germany}

\date{July 18, 2013}

\begin{abstract}
We study charge transport through a topological superconductor with a pair of Majorana end states, coupled to leads via quantum dots with resonant levels. The non-locality of the Majorana bound states opens the possibility of crossed Andreev reflection with nonlocal shot noise, due to the injection of an electron into one end of the superconductor followed by the emission of a hole at the other end. In the space of energies of the two resonant quantum dot levels, we find a four peaked clover-like pattern for the strength of noise due to crossed Andreev reflection, distinct from the single ellipsoidal peak found in the absence of Majorana bound states. 
\end{abstract}


\maketitle

Majorana bound states (MBSs) are zero-energy fermionic states which are their own anti-particles. Since quasi-particles (QPs) in superconductors (SCs) are always superpositions of electron and hole components, the Majorana criterion can be realized in a peculiar way: a zero-energy QP in a SC has equal contributions from electrons and holes, and hence an exchange of electron and hole components leaves the QP invariant. There is currently much interest in the physics of MBSs \cite{MZ2012,WB2012,RL2012,DY2012,DR2012,B2012,A2012,LF2012b}, since one pair of MBSs nonlocally encodes a qubit, which is the building block for fault-tolerant topological quantum computing architectures \cite{Kitaev03,Alicea+11}. 

There is a variety of candidate systems for realizing Majorana fermions. Early proposals considered time-reversal symmetry broken $p$-wave SCs with the candidate Sr$_2$RuO$_4$ \cite{SN2006}. Recently, the SC proximity effect has been suggested as a way to effectively induce $p$-wave pairing in topological insulators \cite{FuKane08} and semiconductors with strong Rashba spin-orbit coupling \cite{SL2010,A2010,LS2010,OR2010}. Recent experiments reported evidence of MBSs in semiconductor-superconductor heterostructures \cite{MZ2012,WB2012,RL2012,DY2012,DR2012}. A possible probe for the nonlocal nature of MBSs is crossed Andreev reflection (CAR), the conversion of an incoming electron into an outgoing hole in a different lead \cite{BF1995,HK1996,M1996,LM2001,RS2001,RC2012}, in contrast to local Andreev reflection (LAR), where electron and hole reside in the same lead. It has been shown theoretically that at sufficiently low voltages and small level width, CAR by the pair of MBS dominates transport \cite{NA2008,BD2006,LL2009,WC2012,GH2011,SB2011,CQ2011}. For voltages larger than the MBS energy splitting $\epsilon_M$ however, resonant tunneling  of electrons and holes gives rise to negative cross-correlations, and the total crossed noise vanishes. 

\begin{figure}[t]
\includegraphics[width=.45\textwidth]{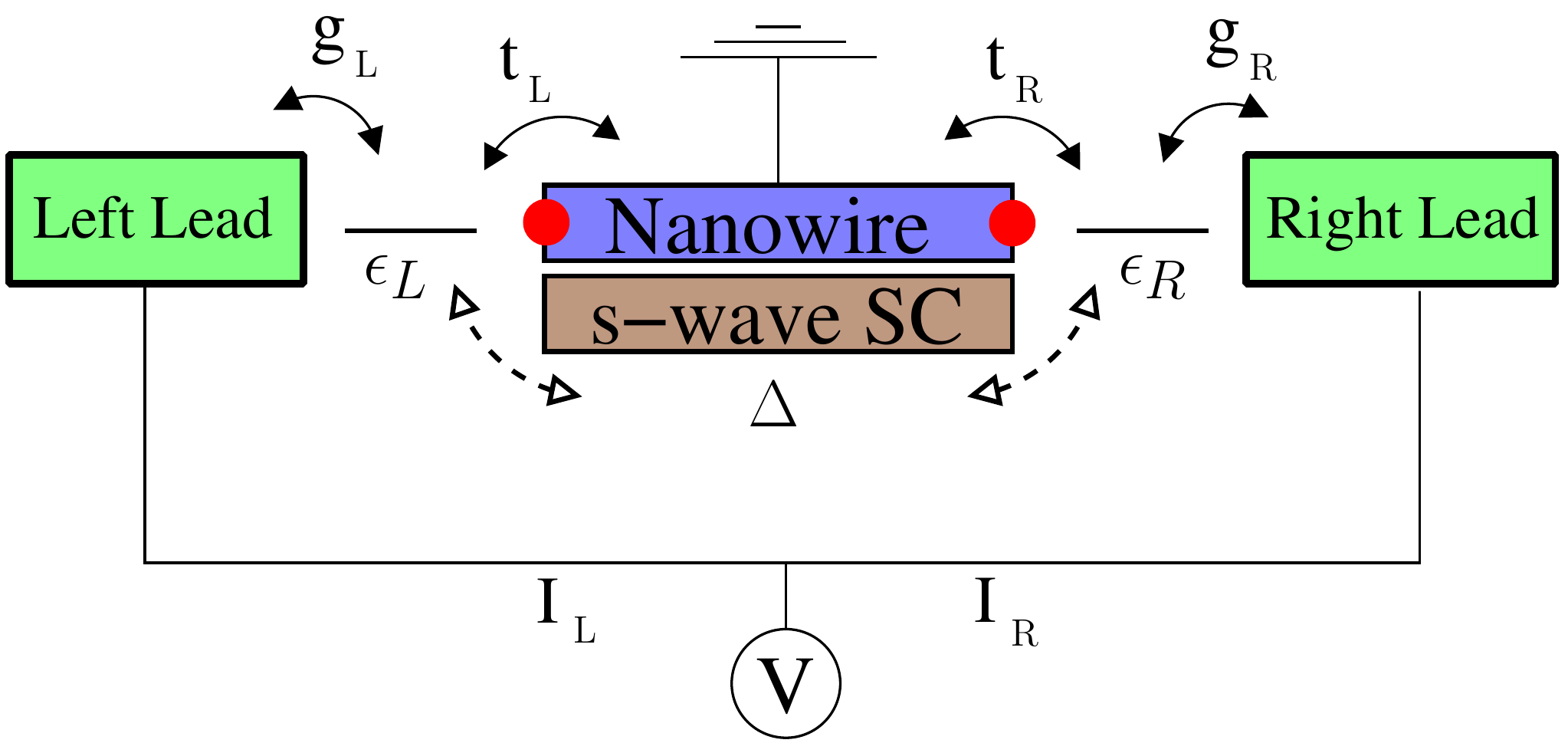}
\caption{(color online) Schematic setup for a system with a pair of Majorana bound states (red dots) coupled to quantum dots which themselves are coupled to lead electrodes. The leads are biased with the positive chemical potential $eV$. Crossed Andreev reflection can be detected by correlating the currents $I_L$ and $I_R$ that flow into the SC nanowire via MBSs. The nearby $s$-wave SC also induces a proximity pairing $\Delta$ between the dots.}
\label{fig:setup2}
\end{figure}

In this letter, we focus on the physics of coupling a pair of MBS at the ends of a wire to leads via resonant quantum dot (QD) levels in the Coulomb blockade regime, see Fig.~\ref{fig:setup2}. As demonstrated in recent experiments \cite{HC2009,HP2010,DR2012b}, the QDs suppress LAR. Due to the finite wire length, the MBSs are tunnel coupled to each other and have an energy splitting $\epsilon_M \sim \Delta \mathrm{exp}(-L/\xi_{SM})$, where $\xi_{SM}$ is the coherence length in the semiconductor. Whenever one of the dot levels is aligned with the chemical potential of the superconductor, an MBS forms on that dot at exactly zero energy \cite{LF2012}, even for $\epsilon_M$ finite. Hence, the MBSs at the ends of the wire are effectively uncoupled, and no CAR can be observed. When tuning the dot levels away from the chemical potential of the superconductor, the coupling between MBSs is restored. In addition, negative cross-correlations due to resonant tunneling are suppressed, and CAR becomes visible in positive current cross-correlations. Thus, the crossed current correlator provides a clear signature of non-local transport through a pair of MBS in the form of a four-leaf clover feature as a function of $\epsilon_L$ and $\epsilon_R$, observable best in the regime of of level broadenings $\Gamma_L, \Gamma_R \gg \epsilon_M$. These findings are in excellent agreement with results for a microscopic model of a spinless $p$-wave SC \cite{K2001}, persist in a more realistic model with several transverse channels, and are robust against addition of disorder. We stress that the mechanism leading to cross-correlations $\propto (e^2/h) \epsilon_M^2/\Gamma$ is a finite energy splitting $\epsilon_M$, and not phase coherent electron teleportation as discussed in \cite{Fu10}. We note that the crossed noise in a similar system was recently studied in Ref. \cite{LS2012} within the diagonalized Master equation approach. There, it was found that the crossed noise stays finite in the limit $\epsilon_M \rightarrow 0$, different from our finding that it is proportional to $\epsilon_M^2$ and thus vanishes. For a discussion of reasons for this disagreement see \cite{supplemental}.

{\it Model system}.--- We consider the Hamiltonians 
%
\begin{subequations}
\begin{align}
H_D&=\sum_{i=L,R}\Big( \epsilon_id_i^\dagger d_i+ g_i d_i^\dagger \psi_i+g_i^* \psi_i^\dagger d_i \Big), \\
H_M&=\epsilon_M i\gamma_L \gamma_R+\sum_{i=L,R} \big( t_i^* d_i^\dagger \gamma_i+t_i \gamma_i d_i \big) ,\\
H_S&= \Delta\big(d_L^\dagger d_R^\dagger +d_Rd_L\big).
\label{eqn:Ham}
\end{align}
\end{subequations}
%
Here, $H_D$ describes two QDs coupled to leads, where $d_i$ annihilates an electron with energy $\epsilon_i$ on dot $i$, $\psi_i$ annihilates a lead electron, and $g_i$ is the lead-dot coupling strength. The lead electrons are characterized by their density of states $\rho_i$, which is assumed to be energy independent, and have a chemical potential $e V$. We consider the regime where the QD single particle level spacing$\delta \epsilon$ satisfies $\delta \epsilon > eV > k_B T$. We assume that the spin degeneracy is lifted by an external magnetic field, and that the QD ground state has an even number of electrons. Then, Kondo physics is absent, and in the Coulomb blockade regime inclusion of only a single dot level in $H_D$ is justified. $H_M$ describes two MBSs with an energy splitting $\epsilon_M$ coupled to the dots. The MBSs are described by hermitian operators $\gamma_i = \gamma_i^\dagger$, which have anti-commutators $\{\gamma_i , \gamma_j\} = 2 \delta_{i,j}$, and are coupled to QD $i$ with amplitude $t_i$. The chemical potential of the SC wire hosting the MBS is zero. $H_S$ describes an additional proximity induced pairing between the dots with an amplitude $\Delta \sim \gamma_S \sin(k_FL)\mathrm{exp}(-L/ \xi_{SC})/(k_FL)$ \cite{RS2001}, where $\gamma_S $ is the normal-state QD level broadening due to the coupling between SC and QD, $k_F$ the Fermi momentum, $L$ the length, and $\xi_{SC}$ the coherence length of the SC. We have in mind that this term may mainly be due to a coupling between the dots and the $s$-wave SC in a hybrid structure. 

We diagonalize the Hamiltonian for MBSs and QDs without lead coupling by solving the corresponding Bogoliubov-de Gennes equation $h \Psi= \epsilon (\mathbb{I}_D + \frac{1}{2} \mathbb{I}_M) \Psi $ with 
\begin{equation}
h=\begin{pmatrix}
0 & i\epsilon_M & t_L & 0 & -t_L^* & 0 \\
 -i\epsilon_M &0 & 0 & t_R & 0 & -t_R^* \\
 t_L^* & 0 & \epsilon_L & 0 & 0 & \Delta \\
 0 & t_R^* & 0 & \epsilon_R & -\Delta & 0 \\
 -t_L & 0 & 0 & -\Delta & -\epsilon_L & 0 \\
 0 & -t_R & \Delta & 0 & 0 & - \epsilon_R 
\end{pmatrix}
\label{eqn:BdG}
\end{equation}
in the basis $\{\gamma_L,\gamma_R,d_L^\dagger,d_R^\dagger,d_L,d_R\}$. Here, $\mathbb{I}_D$ $(\mathbb{I}_M)$ denote the identity matrix in the dot (Majorana) space. In the case $\Delta=0$, the QP energy spectrum has levels at $2\epsilon_M$, $\epsilon_R$, and $\epsilon_L$, with avoided crossings where these levels intersect each other. If one of the dot levels resides at the chemical potential of the SC, e.g. $\epsilon_L=0$, we always find one zero-energy state described by the Majorana operators
\begin{align}
\gamma_1 &=\frac{t_L^* d_L^\dagger + t_L d_L}{|t_L|},\\
\gamma_2 &=\frac{ 2|t_L|(t_R^* d_R^\dagger + t_R d_R-\epsilon_R \gamma_R)+i\frac{\epsilon_M \epsilon_R}{|t_L|} \big(t_L^* d_L^\dagger - t_L d_L \big)} {\sqrt{\epsilon_R^2 \epsilon_M^2  +2|t_L|^2(\epsilon_R^2+2|t_R|^2)}}.
\label{eqn:PMBS}
\end{align}
Here, $\gamma_1$ is localized on the resonant dot, while $\gamma_2$ is partially delocalized, and the weight of $\gamma_2$ on the resonant dot is determined by the energy $\epsilon_R$ of the non-resonant level. In particular for $\epsilon_L=\epsilon_R=0$, we find $\gamma_2= (t_R^* d_R^\dagger + t_R d_R)/|t_R| $ \cite{footnote}. These induced zero-energy states are topologically not protected and acquire a finite energy $\epsilon_L\epsilon_R\epsilon_M/2|t_Lt_R|$ for $\epsilon_L \epsilon_R \neq 0$.

To compute the zero-frequency noise through the above normal-SC-normal (NSN) system, we use a scattering matrix approach which also allows for Andreev reflection processes \cite{AD1996}. This yields the current and the noise correlators 
\begin{align}
I_i &=\frac{e}{h} \int d\epsilon \sum_{\alpha} \mathrm{sign}(\alpha) \sum_{k;\gamma} A_{k,k;\gamma,\gamma}^{(i\alpha)} n_{k,\gamma} ,\\
S_{ij} &=\frac{2e^2}{h} \int d\epsilon \sum_{\alpha,\beta} \mathrm{sign}(\alpha \beta) \sum_{k,l;\gamma,\delta} A_{k,l;\gamma,\delta}^{(i\alpha)} A_{l,k;\delta,\gamma}^{(j\beta)} n_{k,\gamma}(1-n_{l,\delta}),
\label{eqn:Datta}
\end{align}
where Greek indices denote electron (e) and hole (h) channels, $\mathrm{sign}(e)=+1$ and $\mathrm{sign}(h)=-1$, Latin indexes denote the left (L) and right (R) lead, and 
\begin{equation}
A_{k,l;\beta,\gamma}^{(i\alpha)}=\delta_{ik}\delta_{il}\delta_{\alpha \beta}\delta_{\alpha \gamma}-s_{i,k}^{\alpha \beta *}s_{i,l}^{\alpha \gamma} \\. 
\end{equation}
The reservoir distribution functions $n_{k,\gamma}$ are Fermi functions with different chemical potentials for the electron and hole bands $n_{k, \gamma }=1/\big(1+\exp(\beta(\epsilon-\mathrm{sign}(\gamma)eV_k))$. For the setup Fig.~\ref{fig:setup2}, $V_L = V_R \equiv V$. The coefficients $s_{i,j}^{\alpha,\beta}$ are the elements of the $S$-matrix 
\begin{equation}
S(\epsilon)=1 - 2\pi i W^\dagger\left[ \epsilon \, \mathbb{I}_D + \frac{\epsilon}{2} \mathbb{I}_M - h 
+i\pi WW^\dagger\right]^{-1} W,
\label{eqn:smatrix}
\end{equation}
where $W$ describes the coupling between the states of the system without leads and the scattering states in the leads, and $ [\epsilon(\mathbb{I}_D + \frac{1}{2} \mathbb{I}_M) - h +i\pi WW^\dagger]^{-1} $ is the retarded electron Green function for the closed system with self-energy $i\pi WW^\dagger$. The coupling matrix $W$ in the lead basis $\{ \psi^\dagger_L, \psi^\dagger_R, \psi_L, \psi_R \}$ is given by
\begin{equation}
W_{i_{\rm l} \alpha_{i_{\rm l}}, i_{\rm d} \alpha_{i_{\rm d}}} \ = \ \mathrm{sign}(\alpha_{i_{\rm d}}) g_{i_{\rm l}} \sqrt{\rho_{i_{\rm l}}}\delta_{i_{\rm l},i_{\rm d}}\delta_{\alpha_{i_{\rm l}},\alpha_{i_{\rm d}}} \ \ ,
\end{equation}
where $\alpha_{i_{\rm d}}$ $(\alpha_{i_{\rm l}})$ denotes the particle species of QD $i_{\rm d}$ (lead $i_{\rm l}$). The coupling strengths $g_i$ give rise to the level broadening $\Gamma_i=2\pi \rho_i |g_i|^2$ in the dots. In the following, we consider the case $\Gamma_L=\Gamma_R \equiv\Gamma$, $t_L=t_R \equiv t$, and take the limit of zero temperature.

\begin{figure}[tb]
\subfigure{\includegraphics[width=.5\textwidth]{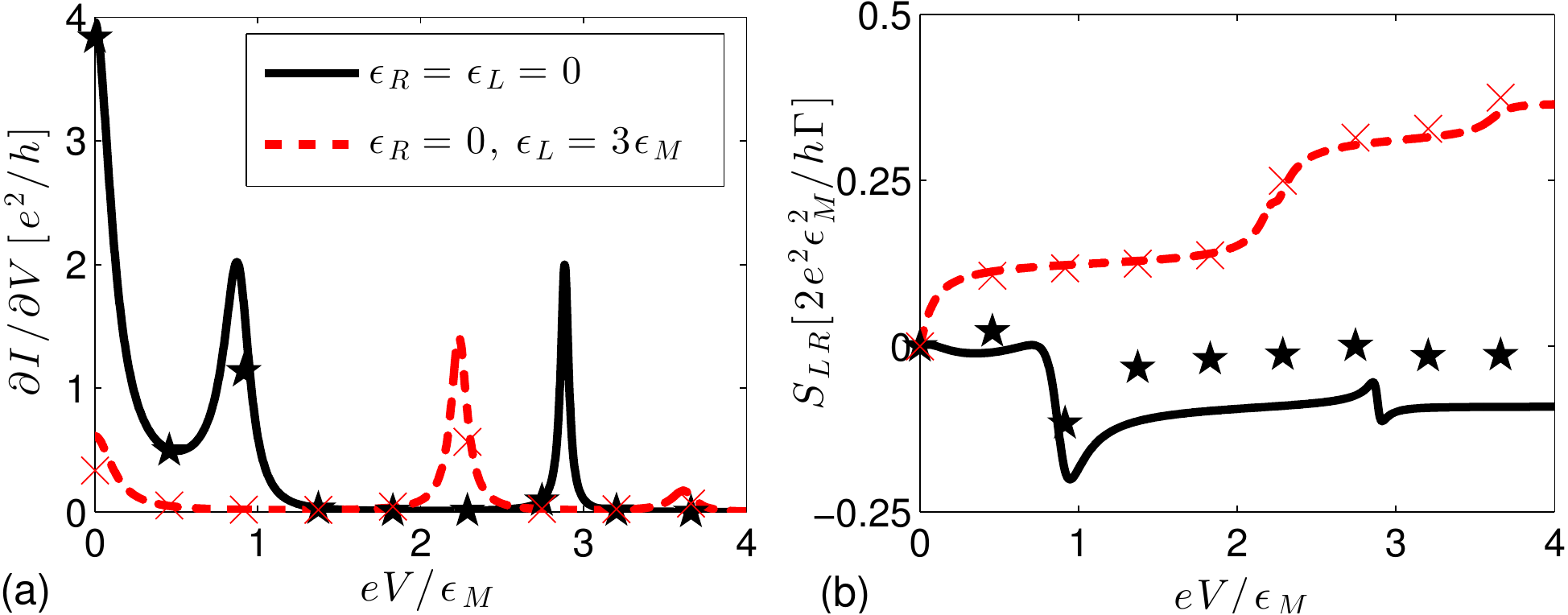}}
\subfigure{\includegraphics[width=.5\textwidth]{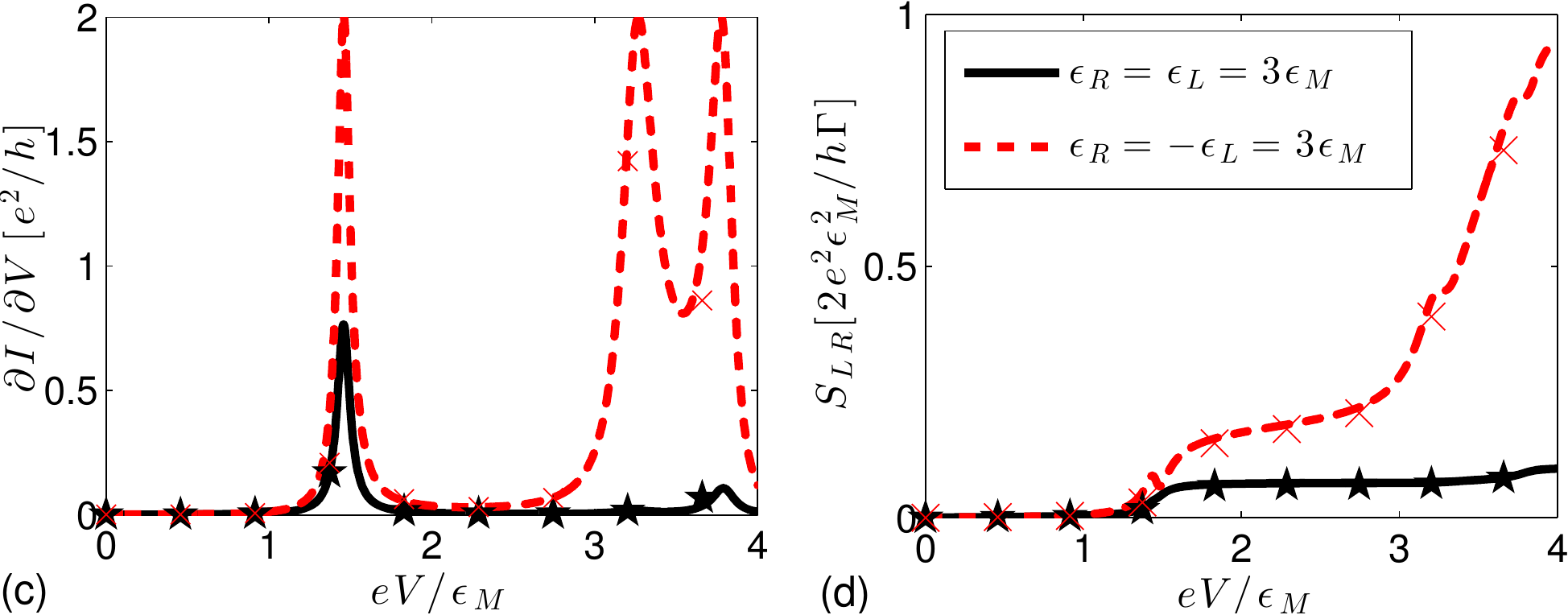}}
\caption{(color online) Current cross-correlator $S_{LR}$ in the weak dot-lead coupling regime with $\Gamma=\epsilon_M/4$, $t=0.8 \epsilon_M$, and $\Delta=0$. The lines for panel (b) are defined in (a), those for panel (c) in (d). The markers denote the results for the spinless SC model with $\epsilon_M=0.01$ meV and $\Gamma=0.002$ meV. }
\label{fig:S12_MBS_low}
\end{figure}

{\it Weak dot-lead coupling}.--- We begin our analysis in the regime $\Delta=0$ and $\Gamma < t<\epsilon_M$. In Fig.~\ref{fig:S12_MBS_low}, both differential conductance and crossed current correlator $S_{LR}$ are displayed as a function of bias voltage for several characteristic points in the $\epsilon_L$-$\epsilon_R$-plane. The differential conductance is peaked at the eigenenergies of Eq.~\eqref{eqn:BdG}. The peak width is determined by the broadening $\Gamma$. If one of the dot levels resides at the chemical potential of the SC, we always find a zero bias peak with height $4 (e^2/h) /[1+\epsilon_M^2(\epsilon_R^2+\Gamma^2/4)/4|t_Lt_R|^2]$ in the differential conductance due to the existence of the induced Majorana states Eq. \eqref{eqn:PMBS}. Since the existence of a zero-energy MBS implies a strongly reduced coupling between left and right side of the wire, we find that these resonances yield only a small contribution to the crossed noise despite their large conductance. 

%
%

In contrast, we do not find a zero-bias conductance peak if both dots are non-resonant. In this regime, there is a striking difference between symmetric ($\epsilon_L=\epsilon_R$) and anti-symmetric ($\epsilon_L=-\epsilon_R$) positions of the dot levels. In both cases, we find contributions to the conductance and $S_{LR}$ due to the hybridization between the dots and the MBS. However, in the anti-symmetric case both the conductance and $S_{LR}$ are much larger than in the symmetric case, and additional resonances at the QD energies contribute to crossed noise. This is due to the fact that Cooper pairs have zero energy, which leads to a suppression of transmission through two resonant levels which have both the same energy in the symmetric case, but allows passage through QDs with opposite level energies in the anti-symmetric case. 

These findings agree very well with results for the microscopic model of a spinless $p$-wave SC defined in Eq.~\eqref{eqn:pwave}, see Fig.~\ref{fig:S12_MBS_low}. The only small deviation in $S_{LR}$ can be seen if both dots are resonant, where the effective model has a small negative $S_{LR}$ for large bias voltage, while it approaches zero for the microscopic model. This deviation has its origin in the presence of an additional transport channel due to a proximity coupling $\Delta$ in the microscopic model, which in principle could be described by the Hamiltonian $H_S$ in Eq.~(1c), but which is not included in the effective model $H=H_M+H_D$ considered here.

\begin{figure}[b]
\subfigure{\includegraphics[width=.5\textwidth]{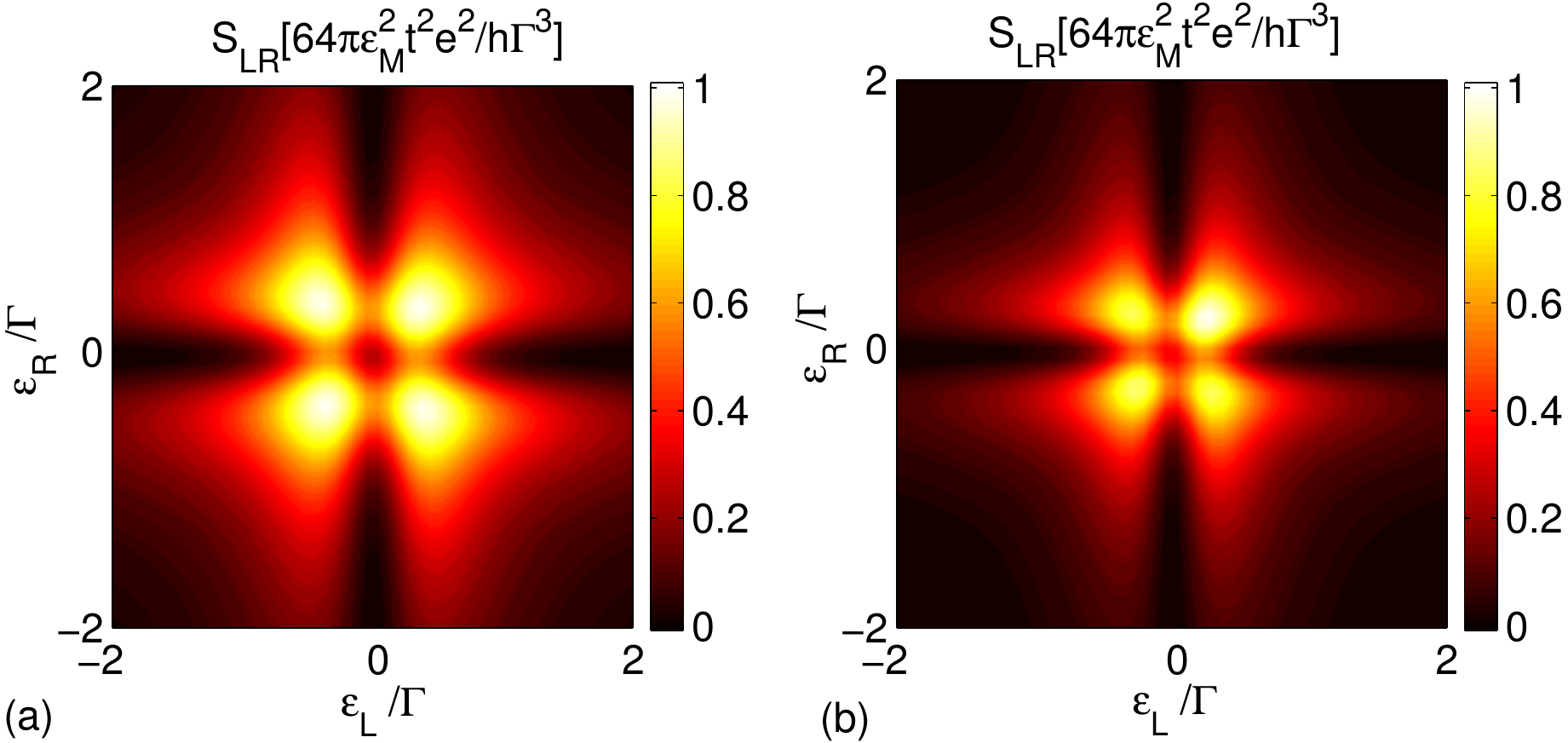}}
\caption{(color online) Current cross-correlator $S_{LR}$ for strong dot-lead coupling. (a) Effective model with $eV=\Gamma/2$, $t=\Gamma/20$, $\epsilon_M= \Gamma/10$, and $\Delta=0$. (b) Spinless SC with $\epsilon_M=0.01$ meV and $\Gamma=0.06$ meV. For both (a) and (b), the pattern changes little for larger $e V$. }
\label{fig:S12_MBS_large}
\end{figure}

{\it Strong dot-lead coupling}.--- We consider the case $t<\epsilon_M \ll \Gamma$ and begin with the situation $\Delta=0$. In Fig.~\ref{fig:S12_MBS_large}(a), the correlator $S_{LR}$ for $\epsilon_M \ll eV=\Gamma/2$ in the $\epsilon_L$-$\epsilon_R$-plane is shown. It is characterized by a four-leaf clover feature with a suppression of crossed noise along lines with either $\epsilon_L=0$ or $\epsilon_R=0$, and peaks at $|\epsilon_L|=|\epsilon_R| \approx \Gamma/2 $. While the peak height scales with $\epsilon_M^2/\Gamma$, the width of these peaks is larger than the Majorana energy splitting due to the large value of $\Gamma$. As before, the suppression of the noise along $\epsilon_L=0$ and $\epsilon_R=0$ is mediated by the formation of zero-energy Majorana modes by virtue of the dot-MBS coupling, which corresponds to the case of uncoupled MBSs. 

%
%

The emergence of an approximate symmetry between symmetric and anti-symmetric positions of the dot levels (absent in the case $\Gamma < t$) can be understood as follows. For large $\Gamma$, the dots are strongly coupled to the leads and effectively become part of them. Hence, there are no separate resonances at the positions of the QD levels anymore, and only a single resonance due to the MBS in the wire survives. Since $t \ll \Gamma$, the broadening of this resonance is much smaller than $\Gamma$. As the QD levels can neither resolve this small broadening of the resonance, nor resolve the location of the resonance, the distinction between symmetric and anti-symmetric QD levels becomes blurred, and the approximate symmetry arises. The Majorana zero-energy state residing on one of the dots for $\epsilon_L=0$ or $\epsilon_R=0$ however does not change its character due to the presence of a large broadening $\Gamma$, and the noise stays low in this case, giving rise to the clover-like pattern in Fig.~\ref{fig:S12_MBS_large}(a). 

In Fig.~\ref{fig:S12_MBS_large}(b), we complement these findings with results for the microscopic model Eq.~\eqref{eqn:pwave}, for which a similar four-leaf clover structure emerges. However, similarly to the weak dot-lead coupling regime, there are small deviations with respect to the effective model near $\epsilon_L=\epsilon_R=0$, mediated by the SC proximity effect.

For finite temperatures $T$, the amplitude of the symmetrically arranged peaks in the clover-like pattern decreases and becomes negative while the anti-symmetrically arranged peaks remain unchanged. Hence, for $k_BT > \Gamma$ the pattern from Fig.~\ref{fig:S12_MBS_large}(a) is modulated in such a way that the peaks for symmetric dot levels become negative of same height \cite{supplemental}.

\begin{figure}[tb]
\subfigure{\includegraphics[width=.5\textwidth]{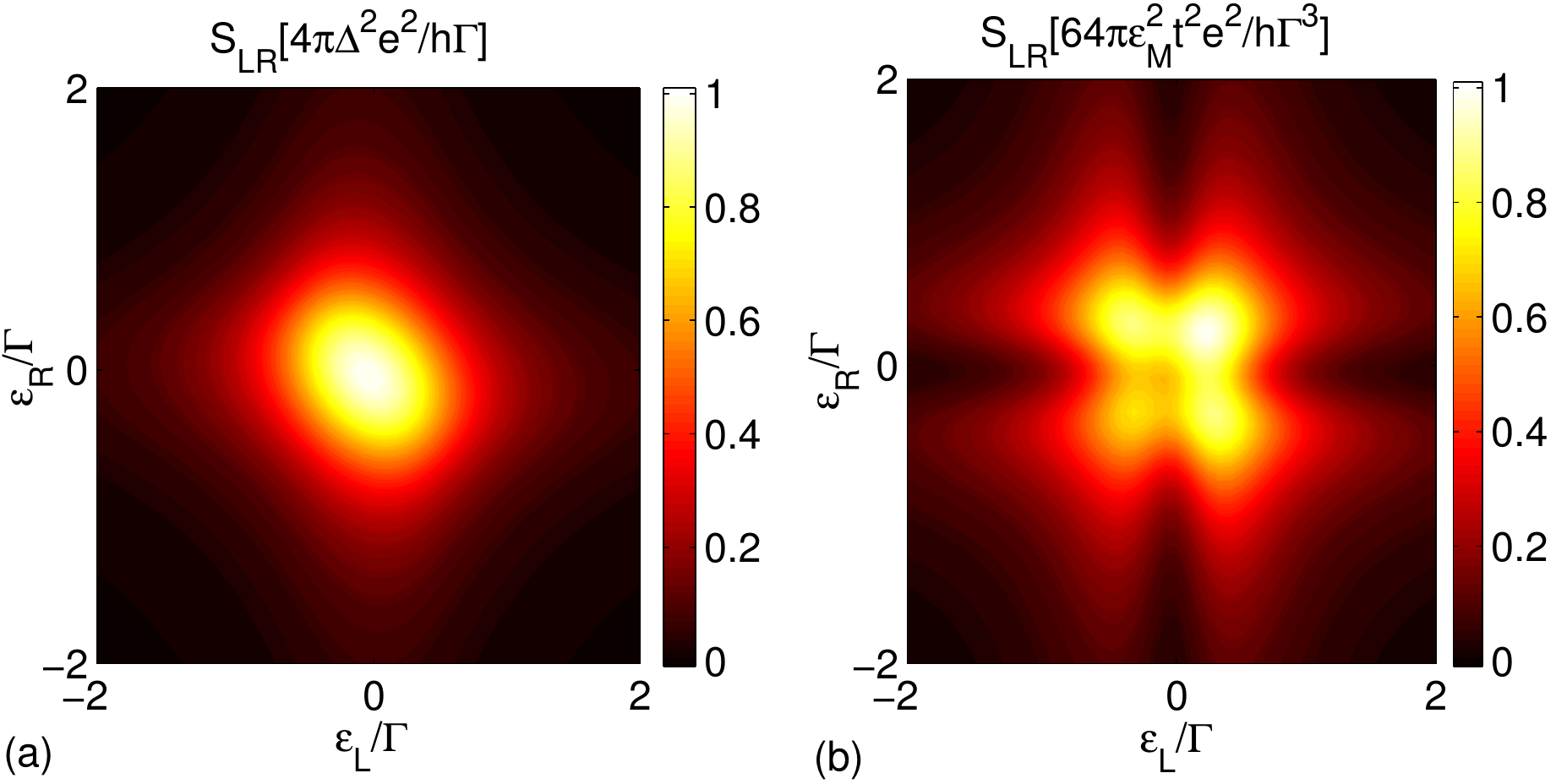}}
\caption{(color online) Current cross-correlator $S_{LR}$ for $eV=\Gamma/2$. The dots are coupled (a) via the SC proximity effect with $\Delta= \Gamma/10$ and (b) via SC proximity effect and with coupling to a pair of MBS with $t=\Gamma/20$, $\Delta=\Gamma/20$, and $\epsilon_M= \Gamma/10$. }
\label{fig:S12_large_swave}
\end{figure}

To gain insight into the effect of an additional proximity term $H_S$, we first discuss the situation without MBS, $H=H_D+H_S$. In Fig.~\ref{fig:S12_large_swave}(a), the crossed current correlator for the SC proximity case is shown. Here, $S_{LR}$ has a single peak of height $\propto \Delta^2/\Gamma$ near $\epsilon_L=\epsilon_R=0$, with width $\Gamma$ along the direction $\epsilon_L=\epsilon_R$, and width $eV$ along the direction $\epsilon_L=-\epsilon_R$. In contrast to the MBS case, there is no additional structure in this peak. 

In figure~\ref{fig:S12_large_swave}(b), we consider the combined Hamiltonian $H=H_M+H_D+H_S$. We find a four-leaf clover feature similar to that in the Majorana only case, with the center of this feature now having a peak due to the proximity term in $H_S$. From this, we conclude that the contributions from the proximity effect and the MBS mediated CAR approximately add up. The relative peak heights in the crossed current correlator reflect the ratio of $\Delta^2 $ and $\epsilon_M^2$. 

%
%

{\it Microscopic model}.--- We complement our calculations by the analysis of a microscopic model for a spinless $p$-wave SC with Hamiltonian \cite{K2001},
\begin{equation}
H_K=-\sum_{j=1}^{N-1} \Big( t_K c_{j+1}^\dagger c_j+\Delta_K c_j c_{j+1} +\mathrm{H.c.}\Big)-\mu_K \sum_{j=1}^N c_j^\dagger c_j,
\label{eqn:pwave}
\end{equation}
where the $c_j$ annihilate a spinless fermion on site $j$ with nearest neighbor hopping $t_K$ and nearest neighbor pairing amplitude $\Delta_K$. This model describes the low-energy physics of a nanowire in the topologically nontrivial phase. In the numerical analysis, we use the parameters $L=1000$ nm for the wire length, $N=200$ sites, $t_K=20$ meV, $\Delta_K=0.8$ meV, and $\mu_K=39.4$ meV, similar to the parameters used in \cite{ZR2011}. These parameter values yield the SC gap $\Delta_{SC}=0.3$ meV and the Majorana energy splitting $\epsilon_M=0.01$ meV. For the coupling of the operators $c_1$ and $c_N$ to the dots, we use $t_{D,K}=0.025$ meV. The results for this model agree very well with those for the effective model Eq.~\eqref{eqn:Ham}, see Figs.~\ref{fig:S12_MBS_low} and \ref{fig:S12_MBS_large}. By introducing a finite wire width, we generalized this model to multichannel $p$-wave SCs where the clover-like pattern remains for transverse channel number $N_\perp < 4\pi \frac{ v_F \Delta_{SC} }{\sqrt{\xi} T_{SD}\Gamma}$ \cite{supplemental}. Here, $v_F$ is the Fermi velocity and $T_{SD}$ the wire-dot coupling strength. For the parameters used in Fig. \ref{fig:S12_MBS_large}(b), this yields the condition $N_\perp  \le 7$. Furthermore, we find that the clover-like pattern is robust against disorder of strength $\lesssim \Delta_{SC}$ \cite{supplemental}. 

The Majorana energy splitting $\epsilon_M$ is oscillating as function of the chemical potential with periodicity $2\pi v_F/L$ when neglecting the long-range Coulomb interaction \cite{SSS2012}. Since the minima of $\epsilon_M(\mu)$ are zero, the Majorana induced current cross correlations vanish. Thus, the chemical potential can be used to switch the crossed noise between the clover-like pattern [Fig. \ref{fig:S12_MBS_large}] and the ellipsoidal pattern [Fig. \ref{fig:S12_large_swave}]. In the experiment this variation of the chemical potential can be realized by applying a global gate voltage to the topologically non-trivial sector of the nanowire.

{\it Conclusion}.--- The non-locality of a pair of Majorana bound states can be probed by crossed Andreev reflection, whose observation is facilitated when suppressing local Andreev reflection with the help of two resonant QDs. In the case of a weak coupling between QDs and leads, we find a set of discrete transmission resonances. When at least one of the QD levels is tuned to the chemical potential of the superconductor, a zero-energy Majorana state forms in the respective QD, which contributes only weakly to crossed Andreev reflection. This feature survives in the limit of strong dot-lead coupling, giving rise to a clover-like modulation of crossed shot noise as a function of QD energies, which is different from the single peak found without Majorana states. 

We acknowledge helpful discussions with A.~Das, M.~Heiblum, and M.~Horsdal, as well as financial support from Federal Ministry of Education and Research (BMBF).

\begin{widetext} 

\section*{Supplementary Material for ``Modulation of Majorana induced current cross-correlations by quantum dots''}

\end{widetext} 

\subsection{Finite Temperatures}

\begin{figure}[htb]
\includegraphics[width=.45\textwidth]{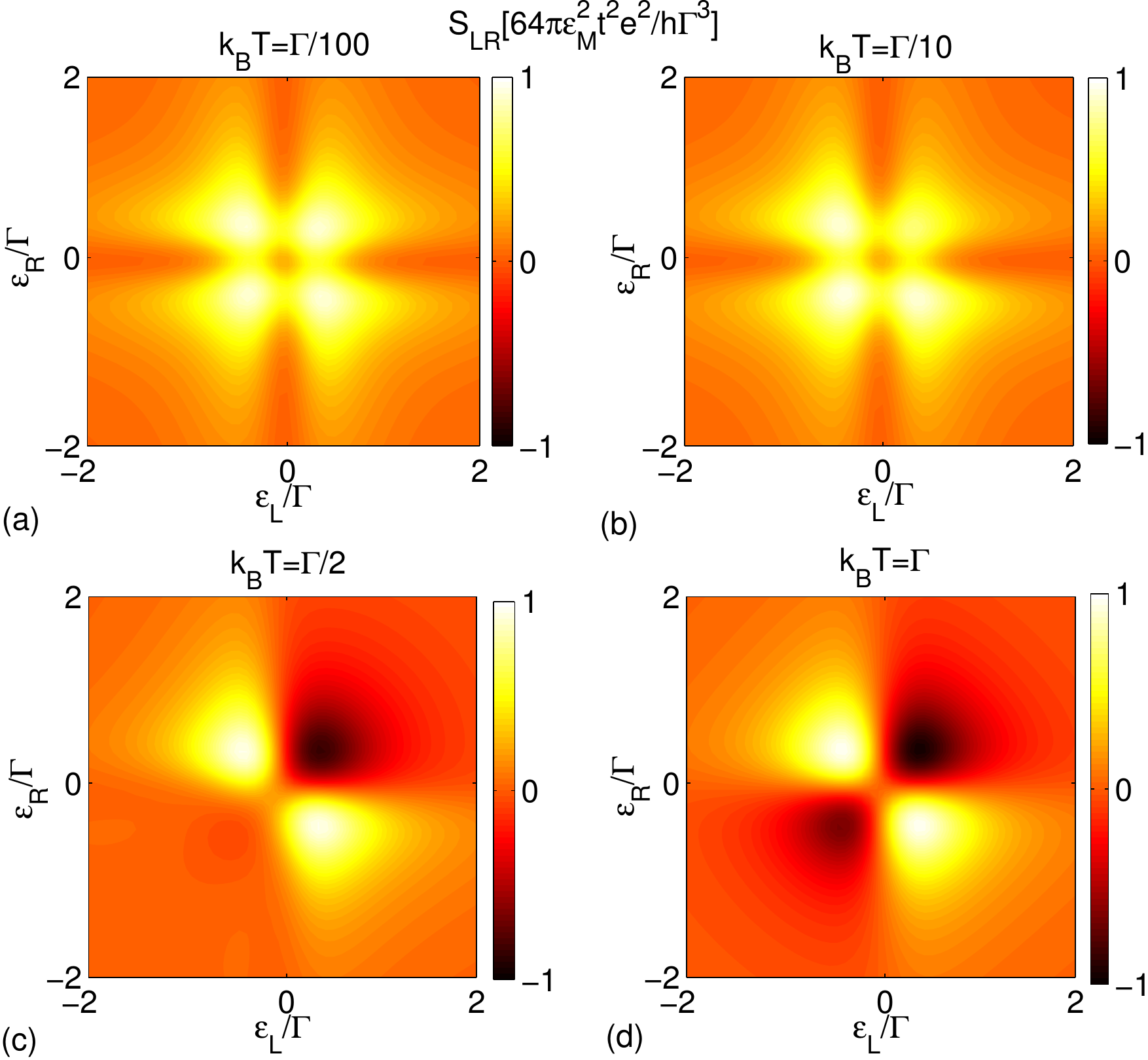}
\caption{(color online) Finite temperature current cross correlations $S_{LR}$ for strong dot-lead coupling with $eV=\Gamma/2$, $t=\Gamma/20$, $\epsilon_M= \Gamma/10$, and $\Delta=0$. (a) $k_BT=0$, (b) $k_BT=\Gamma/10$, (c) $k_BT=\Gamma/2$, and (d) $k_BT=\Gamma$.}
\label{fig:finiteT}
\end{figure}

Here, we study the current cross correlations through the dot-MBS-dot system for finite temperatures $T$. As shown in Fig.~\ref{fig:finiteT}(a), for $k_B T \ll eV$ the crossed noise shows the characteristic four leaf clover-like pattern with equal peak heights. With increasing temperature [Figs.~\ref{fig:finiteT}(b) and (c)] the amplitude of the symmetrically arranged peaks in the clover-like pattern decreases and becomes negative while the height of the anti-symmetrically arranged peaks remains constant. Here, we observe that finite temperatures break the $90^\circ$ rotation symmetry of the cross-noise pattern. For $k_B T \ge eV$, the crossed noise also shows a clover-like pattern, but now with negative height for symmetric and positive height for anti-symmetric dot levels. We attribute the negative cross-correlations for symmetric dot levels $\epsilon_L \approx \epsilon_R$ to resonant tunneling of electrons and holes. For antisymmetric dot levels $\epsilon_L \approx - \epsilon_R$, this resonant tunneling is suppressed and CAR with zero total Cooper pair energy is enhanced, giving rise to positive cross-correlations.

Above we found that for $T=0$ the crossed noise is significantly reduced when at least one of the dot levels lies at the chemical potential of the superconductor. For $k_B T \ge \Gamma$, this suppression becomes complete with vanishing crossed noise along the lines $\epsilon_L=0$ and $\epsilon_R=0$. 

\subsection{Amplitude of the Cross Correlations}

\begin{figure}[htb]
\includegraphics[width=.3\textwidth]{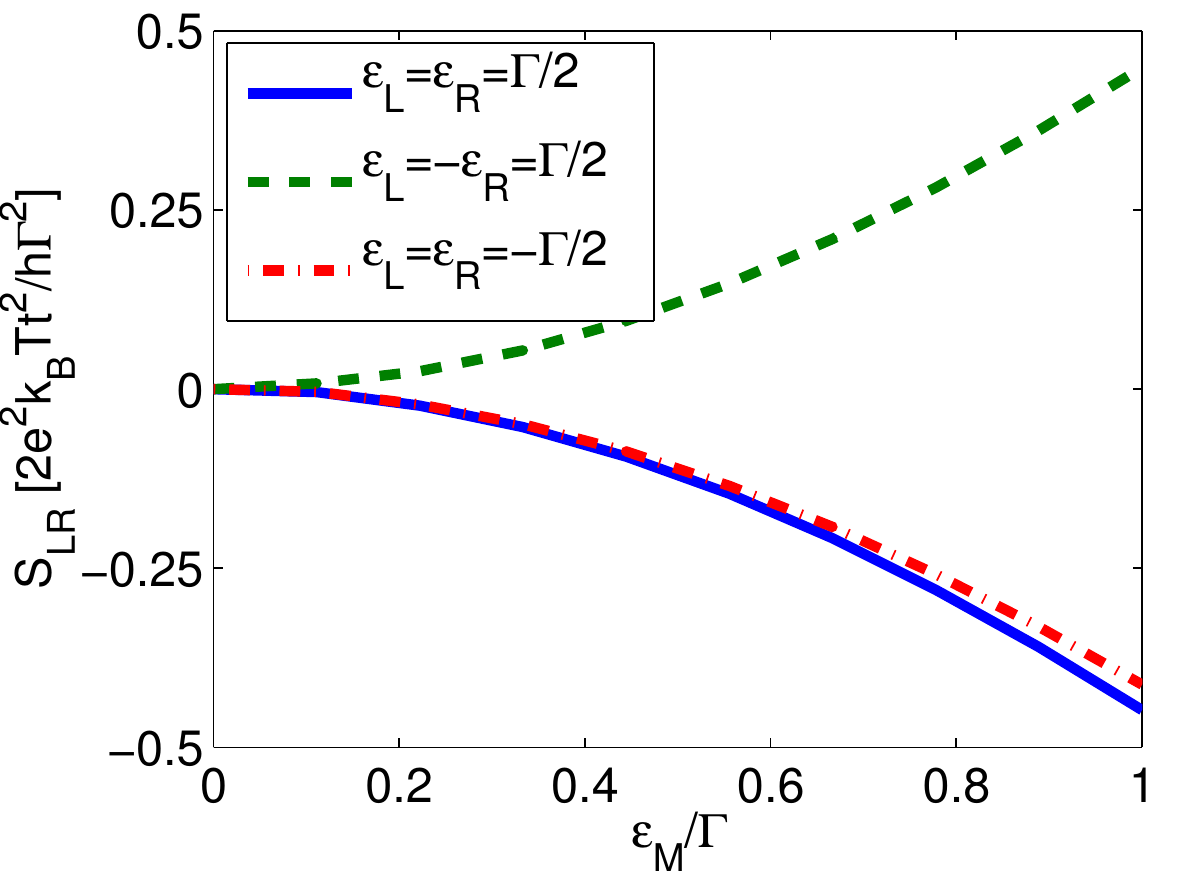}
\caption{(color online) Current cross-correlations $S_{LR}$ for strong dot-lead coupling with $eV=\Gamma/2$, $t=\Gamma/20$, $\epsilon_M= \Gamma/10$, $\Delta=0$, and $T=\Gamma$ as function of the Majorana energy splitting $\epsilon_M$.}
\label{fig:noise_Em}
\end{figure}

For zero temperature, we find the maxima of the clover-like pattern at $|\epsilon_L|=|\epsilon_R|=\Gamma/2$ with amplitude
\begin{equation}
S_{LR,\mathrm{max}}^M=32\pi \frac{2e^2}{h}\frac{\epsilon_M^2t_M^2}{\Gamma^3}
\label{eqn:SLRM}
\end{equation}
with $\epsilon_M,\, t_M\ll \Gamma $. This analytical result is exact for zero temperature. However, for finite temperatures the crossed noise is still proportional to the Majorana energy splitting $\epsilon_M^2$ which is also confirmed by the numerics in Fig.~\ref{fig:noise_Em} where we plot the MBS mediated cross noise as function of $\epsilon_M$. In particular, we find that the cross correlations vanish for vanishing Majorana energy splitting independent of temperature.

For cross correlations induced by the standard superconducting proximity effect, we find for the case $eV/\Gamma \rightarrow \infty$ a maximum along $\epsilon_L+\epsilon_R=0$ with amplitude
\begin{equation}
S_{LR,\mathrm{max}}^S=4\pi \frac{2e^2}{h}\frac{\Delta^2}{\Gamma}
\label{eqn:SLRS}
\end{equation}
with $\Delta \ll \Gamma$. Thus, the current cross correlations vanish for $\Delta=0$ similarly to the MBS case.

\subsection{Transverse Channels}

In the main part of this paper, we investigated the competition between cross correlations mediated by the superconducting proximity effect and MBS. In this section, we study the effect of additional transverse channels $N_\perp$ in the nanowire and estimate a critical channel number for which the cross correlations induced by the proximity effect and the MBS are equal. The coupling Hamiltonian between a multichannel wire and a quantum dot can be written as 
\begin{equation}
H_T=T_{SD} \Big( d^\dagger \psi(\mathbf{r}=0) + \psi^\dagger(\mathbf{r}=0) d \Big),
\label{eqn:coupling}
\end{equation}
where $d$ ($\psi(\mathbf{r}=0)$) denotes the annihilation operator for the dot (wire at site $\mathbf{r}=0$) and $T_{SD}$ the coupling matrix element. We decompose the operator $\psi(\mathbf{r}=0)$ into MBS and delocalized states, $\psi(\mathbf{r}=0)= \gamma_L/\sqrt{2\xi} + \sum_{k,n;E_k>0} \psi_{k,n}/\sqrt{L}$ where $\psi_{k,n}$ denotes the operator for an electron with transverse channel index $n$ and longitudinal momentum $k$. If the energy difference between the subbands is larger than the superconducting gap, we write the coupling Hamiltonian Eq.~\eqref{eqn:coupling} as sum of the coupling between the dot and the MBS and the coupling between the dot and the Bogoliubov quasiparticles,
\begin{equation}
H_T=\frac{T_{SD}}{\sqrt{2\xi}}\big(d^\dagger -d \big)\gamma_L +\frac{T_{SD}}{\sqrt{L}}\sum_{n,k;E_k \ge \Delta_{SC}} \Big( \psi_{n,k}^\dagger d + d^\dagger \psi_{n,k} \Big).
\end{equation}
Hence, we find that the dot-MBS coupling strength is $t_M=T_{SD}/\sqrt{2\xi}$ and the dot-quasiparticle coupling strength is $t_S=T_{SD}/\sqrt{L}$. 

In the previous subsection, we determined the amplitude of the current cross correlations mediated by MBS and the SC proximity effect. Using this result, we find that the relative strength of the Majorana and the proximity induced cross noise is determined by the ratio of Eqs. \eqref{eqn:SLRM} and \eqref{eqn:SLRS}. There, the Majorana energy splitting is $\epsilon_M \approx \Delta_{SC}\sin(k_FL)e^{-L/\xi}/(k_FL)$ and the proximity induced pairing potential is $\Delta \approx \rho_S t_S^2\sin(k_FL)e^{-L/\xi}/(k_FL)$ where $k_F$ denotes the Fermi momentum, $\Delta_{SC}$ the SC gap, and $\rho_S=N_\perp L/2\pi v_F$ the normal state density of states of the nanowire [\onlinecite{RS2001c}]. With $v_F$ denoting the Fermi velocity and $N_\perp$ denoting the number of partially occupied transverse channels, this yields 
\begin{align}
\frac{S_{LR}^M}{S_{LR}^S} =  \Big(\frac{4\pi v_F\Delta_{SC}}{N_\perp  T_{SD}\sqrt{\xi}\Gamma}\Big)^2 . 
\label{eqn:compare}
\end{align}
For observation of the clover like pattern in the current cross-correlations, we demand that the MBS mediated cross noise is larger than the one mediated by the superconducting proximity effect, i.e. $S_{LR}^M>S_{LR}^S$. In this way, we obtain the condition that 
\begin{equation}
N_\perp < 4\pi \frac{ v_F \Delta_{SC} }{\sqrt{\xi} T_{SD}\Gamma}.
\end{equation}

\subsection{Realistic Semiconductor Model}

We consider the Hamiltonian describing a narrow semiconductor nanowire with strong spin-orbit coupling which is predicted to host Majorana bound states~[\onlinecite{SL2010a}-\onlinecite{OR2010a}],
\begin{align}
H=& \int d^2r \Big\{ \sum_{\sigma} \psi^\dagger_\sigma(\mathbf{r})\Big(-\frac{\hbar^2}{2m}\nabla^2 -\mu +E_Z \sigma \Big) \psi_\sigma(\mathbf{r}) \nonumber\\
&+i\alpha \sum_{\sigma,\sigma'} \psi^\dagger_\sigma(\mathbf{r}) \Big( \sigma^y_{\sigma,\sigma'}\frac{\partial}{\partial x}-\sigma^x_{\sigma,\sigma'}\frac{\partial}{\partial y} \Big) \psi_{\sigma'}(\mathbf{r})  \nonumber \\
&+\Delta_{SC}  \Big( \psi^\dagger_\uparrow(\mathbf{r}) \psi^\dagger_\downarrow(\mathbf{r}) + \psi_\downarrow(\mathbf{r}) \psi_\uparrow(\mathbf{r}) \Big)\Big\},
\label{eqn:Hfull}
\end{align}
where $\psi^\dagger_\sigma(\mathbf{r})$ creates an electron at $\mathbf{r}=(x,y)$ with spin $\sigma$, $m$ denotes the effective band mass of the electrons in the nanowire, $\mu$ the chemical potential, $E_Z$ the Zeeman energy due to an external magnetic field, $\alpha$ the Rashba velocity related to the spin-orbit coupling, and $\Delta_{SC}$ the proximity induced $s$-wave pairing potential. 

We use the realistic parameters $\hbar^2/2m = 500 \,\mathrm{meV} \cdot \mathrm{nm}^2$,  $E_Z=1\, \mathrm{meV}$, $\alpha =10 \, \mathrm{meV} \cdot \mathrm{nm}$, and $\Delta_{SC}=0.5 \, \mathrm{meV}$~[\onlinecite{ZHR2011a}]. For certain regimes of the chemical potential, Hamiltonian Eq.~\eqref{eqn:Hfull} can be mapped onto the spinless SC Hamiltonian Eq.~(10) which we use in the main part of this paper. The effective parameters used in the main part are the result of such a mapping, using realistic parameters for InAs and InSb nanowires and $\mu=0$. Without loss of generality, we assume spin polarized quantum dots with spin $\uparrow$ only. In our numerics we consider a nanowire of length $L=520$ nm and width 70 nm. For the above parameters, the superconducting coherence length is $\xi=v_F/\Delta_{\mathrm{eff}}=150$ nm.
\begin{figure}[htb]
\includegraphics[width=.3\textwidth]{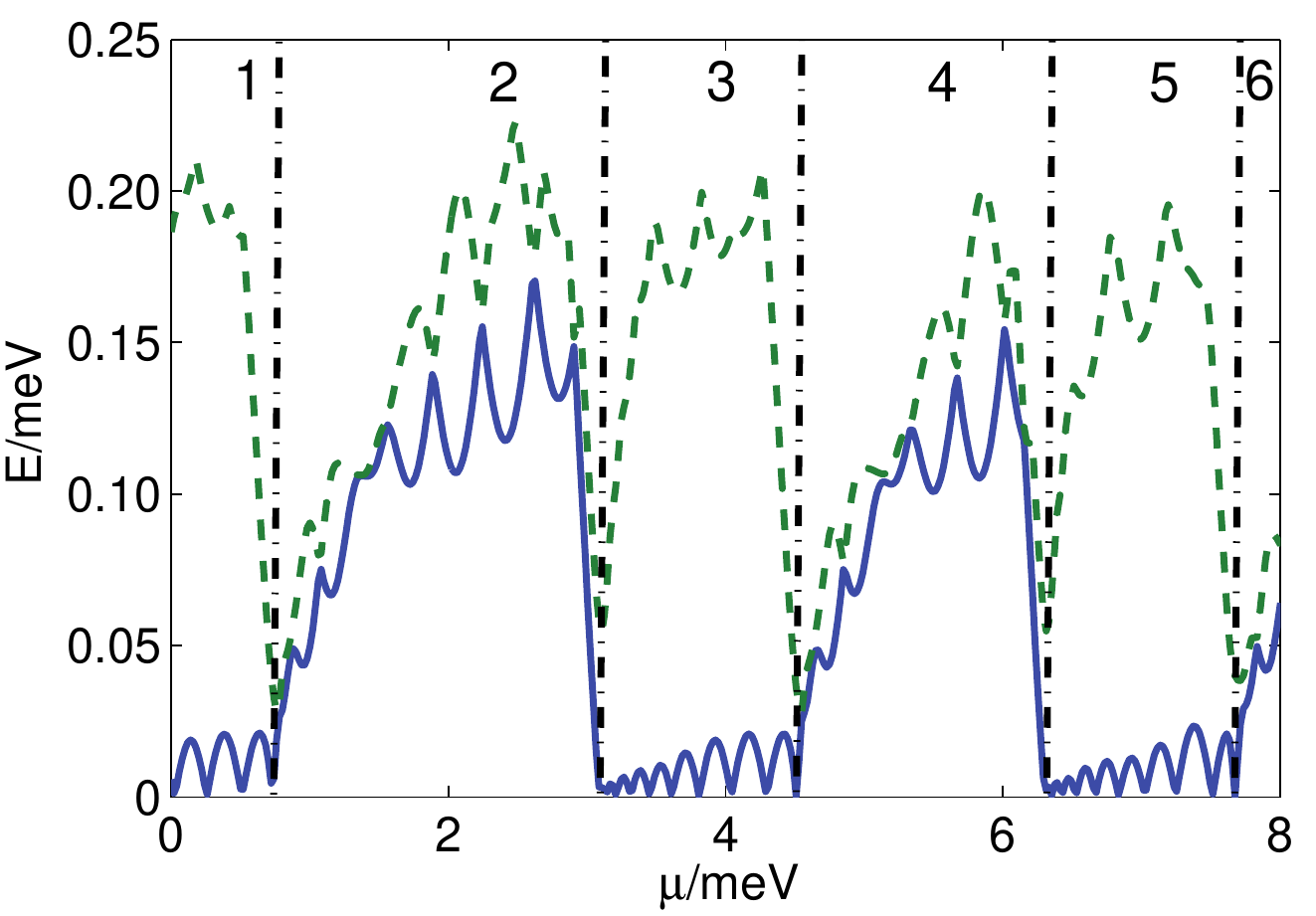}
\caption{(color online) Lowest quasiparticle energy of the semiconductor model as function of chemical potential $\mu$. }
\label{fig:spectrum_mu_lines}
\end{figure}

In the following, we study the current cross correlations for different numbers of transverse channels and investigate the effect of both disorder and small changes of the chemical potential on the clover-like pattern found in the main part of the paper.

\subsubsection{Variation of the Chemical Potential}

In this section, we consider the influence of small changes of the chemical potential on the current cross correlations. In Fig. \ref{fig:spectrum_mu_lines}, we plot the two lowest quasiparticle energies for Hamiltonian Eq.~\eqref{eqn:Hfull}. Here, a topologically non-trivial phase exists in sectors with a sub-gap state of energy $\epsilon_M \ll \Delta$. This low-energy state corresponds to two coupled Majorana bound states with energy splitting $\epsilon_M$. Such a state always exists if an odd number of subbands is partially occupied. As function of the chemical potential the Majorana energy splitting oscillates with period $2\pi v_F/L$ and with energy minima of $\epsilon_M=0$. 

\begin{figure}[htbp]
\includegraphics[width=.45\textwidth]{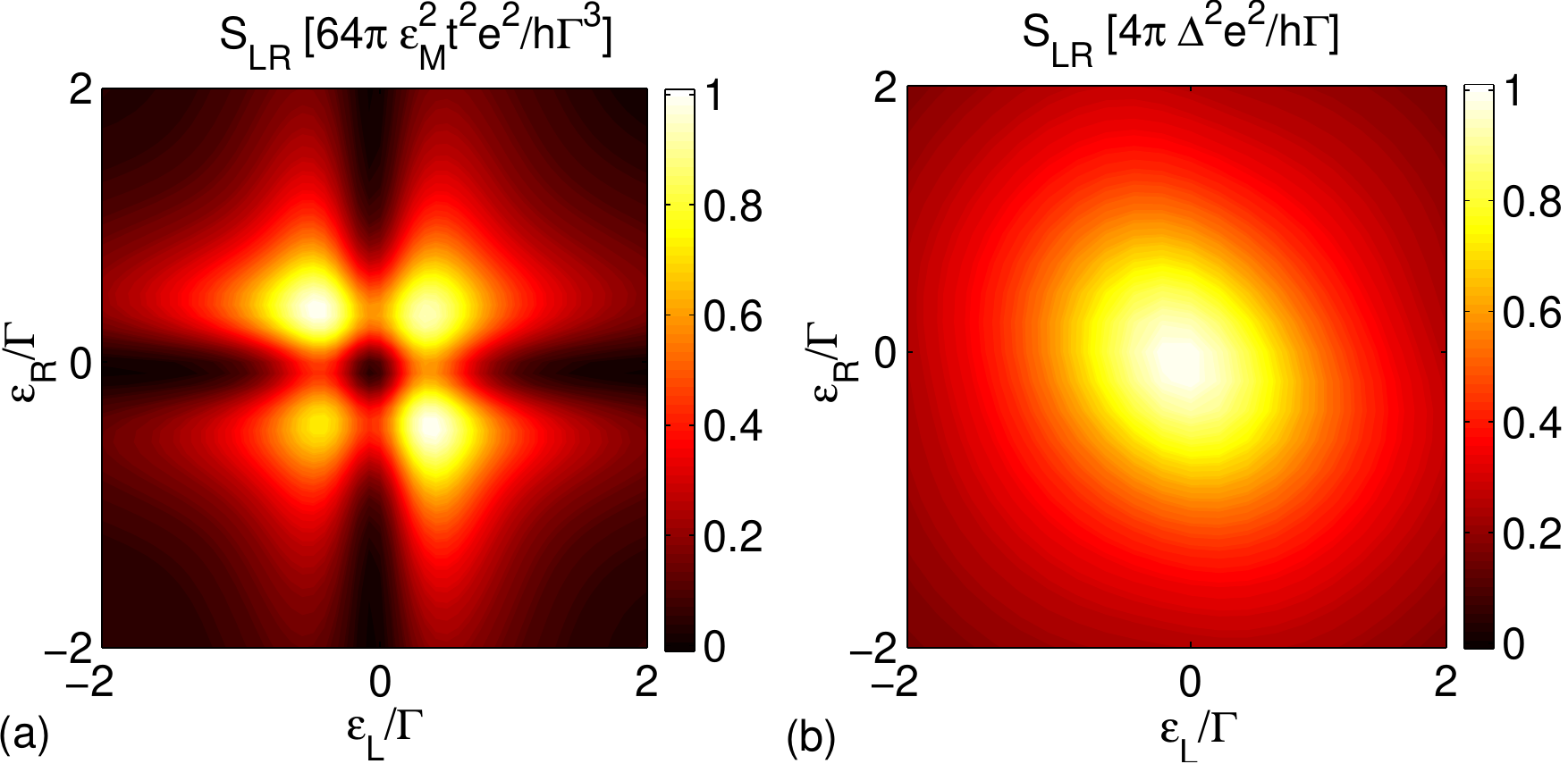}
\caption{(color online) Current cross correlations $S_{LR}$ for $\Gamma=0.04$ meV, and $T_{SD}=0.12$ meV$\cdot$nm$^{1/2}$. (a) Clover-like pattern for $\mu=0.1$ meV with $\epsilon_M=0.01$ meV and (b) elliptic pattern for $\mu=0$ meV with $\epsilon_M=0.001$ meV.  }
\label{fig:oscillation}
\end{figure}

In Fig. \ref{fig:oscillation}, we plot the current cross correlations for two values of the chemical potential for which the Majorana energy splitting has a local maximum and a local minimum. We find that the patterns in the current cross correlations are very different in the two cases, with a clover-like pattern for $\epsilon_M \neq 0$ and an ellipsoidal pattern for $\epsilon_M=0$. This is in full agreement with our findings that the Majorana induced current cross correlations are proportional to $\epsilon_M^2$. Thus, we conclude that small variations of the chemical potential can be used as a tool to switch between different patterns of current cross-correlations. Such a switching mechanism does not exist in the topologically trivial phase and is thus a signature for Majorana bound states with oscillating Majorana energy splitting. The change of the chemical potential can be realized by applying a global gate voltage. An alternative route to demonstrate the oscillations is a change of the magnetic field which gives rise to oscillation of periodicity $\omega_B=4\pi v_F/g\mu_B L$.

Above we have discussed that the proximity induced pairing oscillates as function of $k_FL$ which changes when changing the chemical potential. However, for the standard proximity coupling in semiconductor nanowires we do not find a unique Fermi momentum because of the spin-orbit coupling. Thus the oscillation of the proximity induced pairing potential is smeared out and we always find a non-zero contribution of the superconducting proximity effect to the crossed noise.

\begin{figure}[htb]
\includegraphics[width=.45\textwidth]{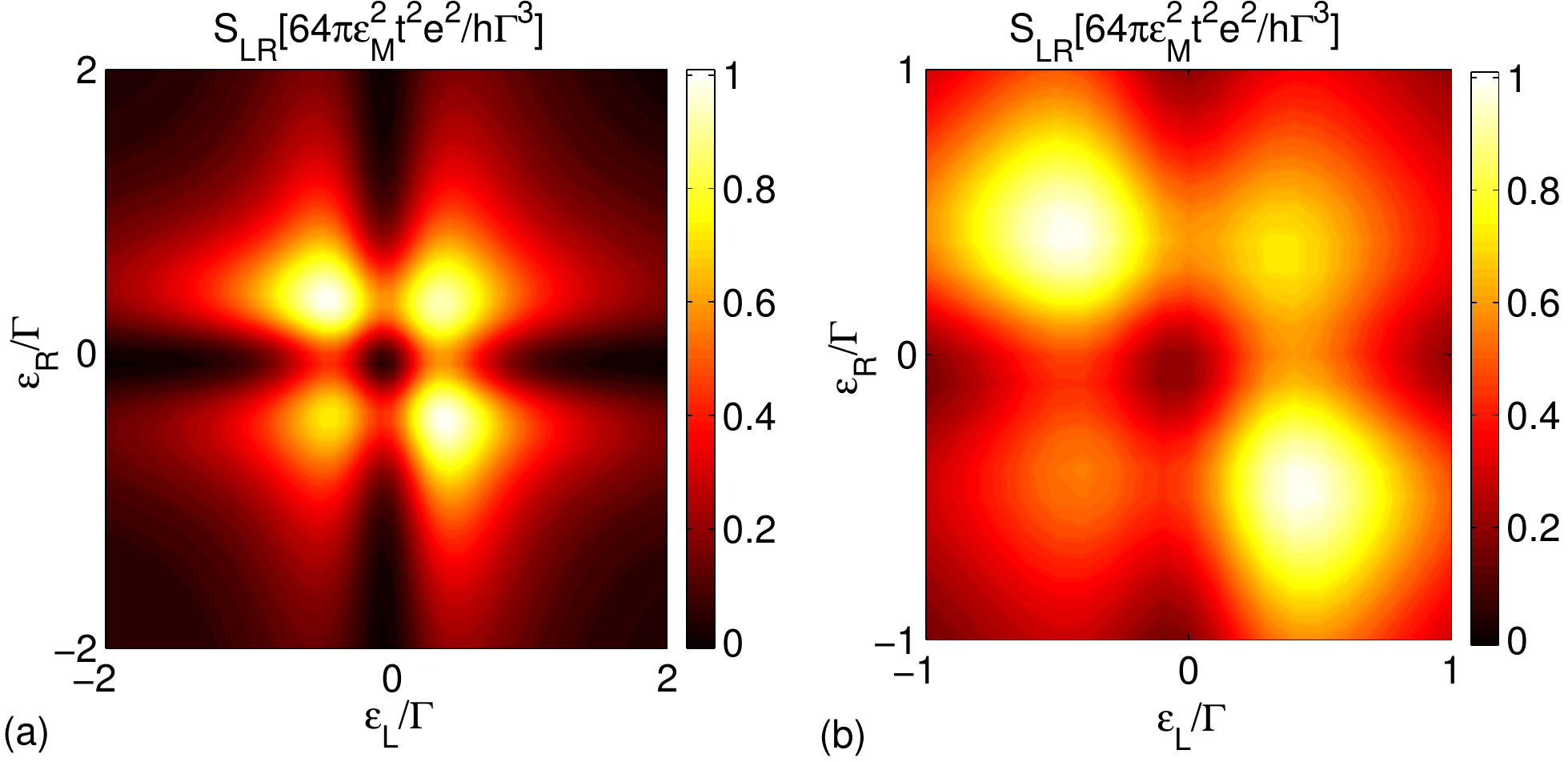}
\caption{(color online) Current cross correlations $S_{LR}$ for $\Gamma=4\epsilon_M$, and $T_{SD}=4\sqrt{10 }\epsilon_M$ nm$^{1/2}$. (a) Three-band semiconductor of width 70 nm with $\mu=4.1$ meV and (b) five-band semiconductor of width 90 nm with $\mu=6.1$ meV.  }
\label{fig:threeband}
\end{figure}

\subsubsection{Multiband Systems}

Due to the finite extension of the nanowire in $y$-direction, we expect to find a multi-band system where the bands have a separation of several meV. For an even number of occupied transverse channels, the wire is in the topologically trivial phase, i.e. the Majorana bound states are absent and thus we find a single ellipsoidal cross-noise pattern similar to the one in Fig. 4 of the main part of this paper where we studied the current cross correlations through a pair of quantum dots with superconducting pairing. For an odd number of occupied transverse channels, the wire is in the topologically non-trivial phase with Majorana end states. In Fig.~\ref{fig:threeband}, current cross correlations are shown for $\mu=4.1$ meV and width 70 nm which corresponds to the three-band case and for $\mu=6.1$ meV and width 90 nm which corresponds to the five-band case. In both case, we still find the characteristic four leaf clover-like pattern in current cross-correlations, similar to the single-band case. However, the amount of noise for $\epsilon_L=\epsilon_R=0$ is increased by a factor of $\approx 9$ in the three-band system as compared to the one-band case, as expected from the estimate Eq.~\eqref{eqn:compare}. This numerical finding confirms our analytical result Eq.~\eqref{eqn:compare} that the clover-like pattern is not restricted to the single-band wire and can also be found in multi-band wires.

\subsubsection{Electrostatic Disorder}

\begin{figure}[htb]
\includegraphics[width=.45\textwidth]{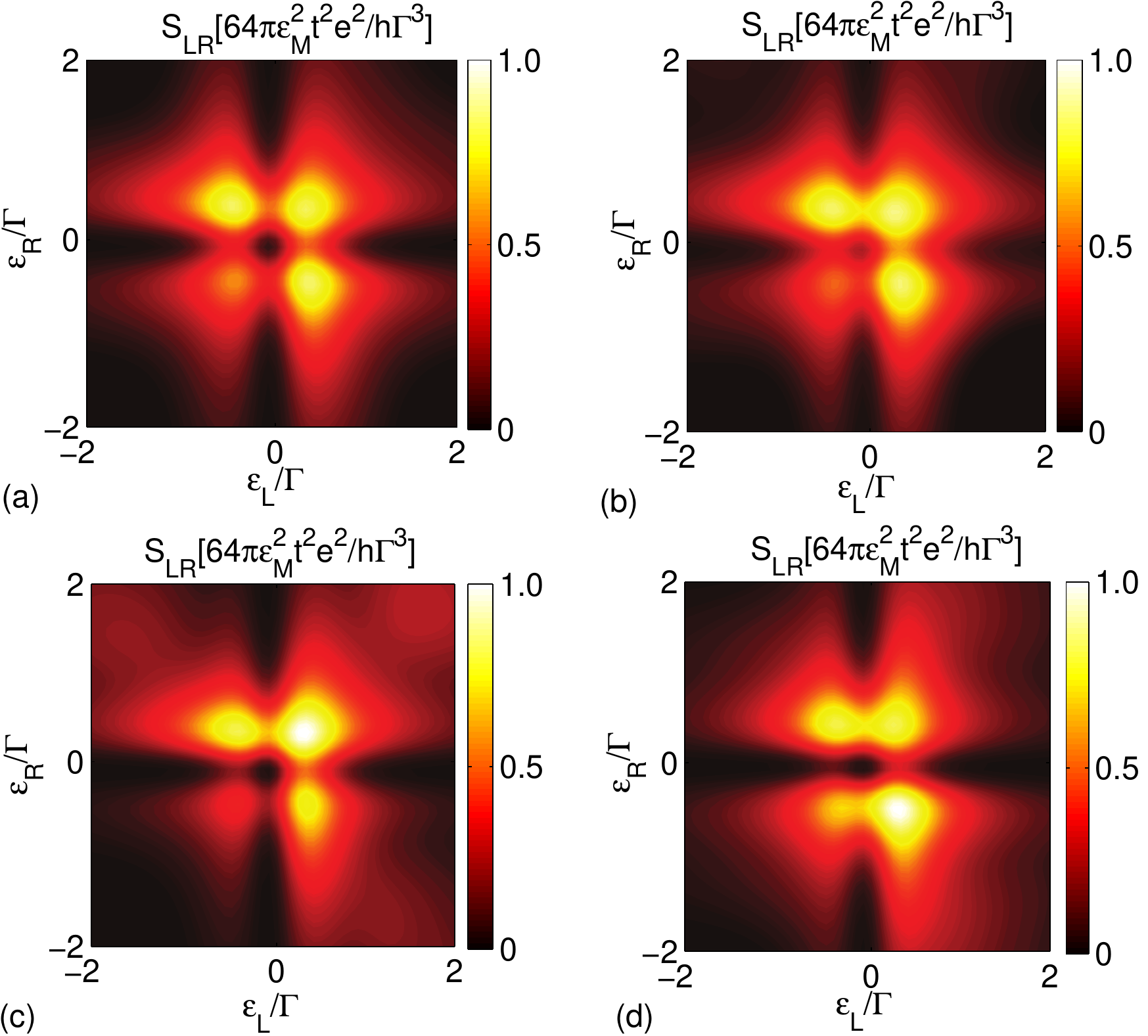}
\caption{(color online) Current cross correlations $S_{LR}$ of the single-band semiconductor model with $\mu_0=0$ for different disorder strengths. Ensemble averaged cross noise for disorder strengths (a) $U=\Delta/4$ and (b) $U=\Delta/2$. Crossed noise for a characteristic disorder realization with (c) $U=\Delta/4$ and (d) $U=\Delta/2$.  }
\label{fig:disorder}
\end{figure}

In this section, we consider a spatially fluctuating chemical potential with mean value $\mu_0$ and random variations $\delta\mu(\mathbf{r})$ with $\langle \delta\mu(\mathbf{r}) \delta\mu(\mathbf{r}') \rangle = U^2 \mathcal{V} \delta(\mathbf{r}-\mathbf{r}') $. In Figs. \ref{fig:disorder}(a) and (b), we display the disorder averaged current cross correlations for disorder strengths $\Delta_{SC}/4$ and $\Delta_{SC}/2$. We here averaged over 50 random disorder configurations, and find that the clover-like pattern is robust with respect to electrostatic disorder. 

In Figs. \ref{fig:disorder}(c) and (d), we display the current cross correlations for single characteristic disorder configurations of strengths $U=\Delta_{SC}/4$ and $U=\Delta_{SC}/2$, respectively. When comparing the cross-correlations for a random configuration with the clean case, we find that electrostatic disorder distorts the clover-like pattern and disorder averaging averages over distortions which restores the clover-like pattern as shown in Figs. \ref{fig:disorder}(a) and (b).

\subsection{Relation to Previous Work}

In the recent work [\onlinecite{LS2012}] (short LLS in the following), current cross-correlations in a setup similar to ours were studied by using the diagonalized master equation approach in the sequential tunneling regime. In particular, for $\epsilon_M=0$ finite current cross-correlations were found, in disagreement with our result that current cross-correlations are proportional to $\epsilon_M^2$ and should thus vanish in the limit $\epsilon_M \to 0$. In the following, we discuss possible reasons for this discrepancy. 

The physical conditions under which the diagonalized master equation approach is justified are (i) that the bath correlation time is small compared to the relaxation time of the dot-MBS-dot system, i.e. weak coupling between the leads and the dot-MBS-dot system $\Gamma \ll k_B T$, and (ii) that the excitation energies $\Delta E$ within each parity sector are large compared to $\Gamma$ [\onlinecite{BP2002}]. For $\epsilon_M=0$, the states with different parity are always degenerate, which should be unproblematic since the coherent superposition between these states is unimportant for electronic transport. However, for energies $\epsilon_M=\epsilon_L=\epsilon_R=0$, LLS find in their Eq.~(16) an additional degeneracy of the two lowest lying states in each sector, which is problematic since it violates condition (ii). Thus, we conclude that for the energy spectrum used by LLS the diagonalized rate equation approach is not appropriate in the vicinity of this point.

Nonetheless, if one forgoes the question of whether the diagonalized master equation approach is applicable,
we can compare the single-particle energy spectrum we find by solving the Bogoliubov-de Gennes (BdG) equations with the many-particle energy spectrum discussed by LLS when diagonalizing the Hamiltonian in the eight-dimensional many-body Fock space. Since parity is a good quantum number for the isolated dot-MBS-dot system, it is possible to decompose the Fock space into two four-dimensional subspaces with even and odd parity, and to diagonalize the Hamiltonian in each subspace separately. Then, the ground state is given by the vector with lowest energy, and the parity changing excitations are described by many-body wave functions with a parity different from that of the ground state. In particular, for $\epsilon_M=\epsilon_L=\epsilon_R=0$ and $|t_L|=|t_R|=t$, LLS find in their Eq.~(16) that the states for even and odd parity are degenerate, and that each sector has energies $\{-\sqrt{2}t,-\sqrt{2}t,\sqrt{2}t,\sqrt{2}t\}$. Thus, the excitation energies for parity changing excitations of the ground state are  $\{0,0,2\sqrt{2}t,2\sqrt{2}t\}$. In contrast, in our manuscript we use the BdG formalism to diagonalize the Hamiltonian. The BdG formalism is a single-particle formalism based on the single-particle Schr\"odinger equation and describes quasiparticle excitations above the ground state. Since the BdG formalism doubles the physical Hilbert space, only three out of the six eigenvalues obtained by diagonalizing the Hamiltonian Eq.~(2) are independent solutions. Excited states can be constructed by adding one quasiparticle (three possible states), two quasiparticles (three possible states), or three quasiparticles (one state). Thus, together with the ground state, these states span an eight-dimensional Fock space, in agreement with LLS. Using the BdG formalism, we find for parameters  $\epsilon_M=\epsilon_L=\epsilon_R=0$ the single-particle excitation energies $\{0,2t,2t\}$, and as a consequence the three-particle excitation energy $4t$. Thus, the energy difference between the many-body ground state and parity changing excited states should be $\{0,2t,2t,4t \}$, different from the excitation spectrum obtained above by using the energies of LLS. This discrepancy in the energy spectrum casts additional doubt on the results of LLS and their interpretation. In addition, even when using the  correct energy spectrum,  there exist degeneracies between excited states within each parity sector for the choice of parameters $|\epsilon_R|=|\epsilon_L|$, where the current cross-correlations are strongest. Therefore,  it seems that the applicability of the diagonalized rate equation approach to the dot-MBS-dot system is limited.


\begin{thebibliography}{99}
\bibitem{footnote}

Formally, we can decompose the QD fermion operators into a pair of Majorana operators each. 
There exists a particular decomposition for which only one of the dot Majoranas couples to the adjacent MBS. In the limit where one 
or both of the dot levels $\epsilon_L, \epsilon_R =0$, the two Majoranas in the dot are not coupled to each other, such that  an uncoupled Majorana resides on the dot.
\end{thebibliography}
\end{document}